\documentclass[conference]{IEEEtran}
\IEEEoverridecommandlockouts
\usepackage{hyperref}
\usepackage{cite}
\usepackage{amsmath,amssymb,amsfonts}
\usepackage{algorithmic}
\usepackage{graphicx}
\usepackage{textcomp}

\usepackage{pgf}

\usepackage{xcolor}
\usepackage{multirow}
\usepackage{dblfloatfix}  
\def\BibTeX{{\rm B\kern-.05em{\sc i\kern-.025em b}\kern-.08em
    T\kern-.1667em\lower.7ex\hbox{E}\kern-.125emX}}
\begin{document}

\title{Multi-Environment based Meta-Learning with CSI Fingerprints for Radio Based Positioning}

\author{\IEEEauthorblockN{Anastasios Foliadis\IEEEauthorrefmark{1}\IEEEauthorrefmark{2}, Mario H. Casta\~{n}eda Garcia\IEEEauthorrefmark{1}, Richard A. Stirling-Gallacher\IEEEauthorrefmark{1},  Reiner S. Thom\"a\IEEEauthorrefmark{2}}
	\IEEEauthorblockA{\IEEEauthorrefmark{1}\textit{Munich Research Center}, \textit{Huawei Technologies Duesseldorf GmbH}, 
		Munich, Germany \\
		\textit{\IEEEauthorrefmark{2}Electronic Measurements and Signal Processing}, \textit{Technische Universit\"at Ilmenau}, Ilmenau, Germany\\
		\{\href{mailto:anastasios.foliadis@huawei.com}{anastasios.foliadis}, 
		\href{mailto:mario.castaneda@huawei.com}{mario.castaneda}, 
		\href{mailto:richard.sg@huawei.com}{richard.sg}\}@huawei.com, 
		\href{mailto:reiner.thomae@tu-ilmenau.de}{reiner.thomae@tu-ilmenau.de}}}

\maketitle

\begin{abstract}

Radio based positioning of a user equipment (UE) based on deep learning (DL) methods using channel state information (CSI) fingerprints have shown promising results. DL models are able to capture complex properties embedded in the CSI about a particular environment and map UE's CSI to the UE's position. However, the CSI fingerprints and the DL models trained on such fingerprints are highly dependent on a particular propagation environment, which generally limits the transfer of knowledge of the DL models from one environment to another. In this paper, we propose a DL model consisting of two parts: the first part aims to learn environment independent features while the second part combines those features depending on the particular environment. To improve transfer learning, we propose a meta learning scheme for training the first part over multiple environments. We show that for positioning in a new environment, initializing a DL model with the meta learned environment independent function achieves higher UE positioning accuracy compared to regular transfer learning from one environment to the new environment, or compared to training the DL model from scratch with only fingerprints from the new environment. Our proposed scheme is able to create an environment independent function which can embed knowledge from multiple environments and more effectively learn from a new environment.


\end{abstract}


\section{Introduction}
\label{sec:intro}

One of the main requirements of future communication networks is accurate user positioning. The precise position information can aid future applications and improve their performance. Accurate positioning in wireless networks is enabled by the support of multiple antennas and large available bandwidths of current and future communication standards. 

Generally, classical positioning methods adopt a 2-step approach. First, some relevant parameters are extracted from the measured channel state information (CSI), e.g. angle of arrival (AoA), time of arrival (ToA), etc. In the second step, the parameters are appropriately combined to obtain a position estimate. The function that extracts the channel parameters can be considered environment independent, while the mapping of the parameters to the position inherently depends on the propagation characteristics of the particular environment.

Recently there has been an increased focus on using deep learning (DL) to aid positioning by leveraging the ability to collect large amounts of data. Specifically, DL based positioning methods can exploit the information that is embedded in a multi-path channel between the user equipment (UE) and the base station (BS), which can be considered a unique fingerprint of the user's location. For instance, a neural network (NN) can be trained on a database of uplink CSI fingerprints of different positions of a UE along with the respective UE's location labels. The NN can later be used for estimating a UE's position, by mapping the CSI of the UE to an estimated UE’s position. 

DL based positioning methods have shown promising results, achieving sub-meter accuracy in some indoor and outdoor environments \cite{Foliadis2021CSIBasedLW}. The main benefit of using DL in comparison to classical positioning approaches is that it can be employed in environments with non line of sight or with severe multipath
where classical positioning methods may be impaired or fail. On the other hand, one of the major downsides of employing DL based positioning methods with fingerprints is that the knowledge that is acquired by training a NN on a particular environment is generally not directly applicable to other environments. In other words, a change in the environment may cause the model to completely fail or significantly reduce its positioning accuracy. The effect that a potential change in the environment has on the NN model's performance is explored in \cite{Foliadis2022ReliableDL}.

The most straightforward way to address a change in the environment is to train a new DL model for the new environment. A more efficient approach though, is to simply fine-tune a previously trained model on fingerprints from the new environment. This process is called transfer learning and it's a way of re-using the knowledge of the initial model for the new environment. The idea behind transfer learning is that the initial layers of a NN learn to extract relevant features from the input data, while the final layers combine those features in a task specific way. By re-training the model, only a subset of its parameters need to be majorly adjusted, which speeds up training, can improve performance and requires a reduced amount of training data. In \cite{Bast2020CSIbasedPI } it is shown that fine-tuning a model using data from a new environment can even outperform a model that is trained from scratch on data only from the new environment. 
Transfer learning across real environments and between simulated and real environments is examined in \cite{Stahlke2022TL5G}.

Motivated by the two step approach of classical positioning methods, we propose a DL model to support transfer learning, that analogously consists of two parts. For the proposed DL model, we aim that the initial layers extract location-related features from the measured CSI from a UE, in a way that it is independent of the environment. The latter layers of the proposed DL model then combine the features in an environment dependent way to obtain the UE's position estimate. To strive to have the initial layers to be independent of the environment, we propose a multi-environment meta learning scheme to train the first part of the DL model with data from multiple different environments. The second part of the DL model would then be trained to be environment specific. This is in contrast to the approach presented in \cite{Klus2021TransferLF}, where the second part of the DL model is trained on different received signal strength (RSS) datasets.
In the context of deep learning the process of applying an initial learning scheme with the goal of improving transfer learning is included under the umbrella term of Meta-Learning \cite{Hospedales2021MetaLearningIN}.

In this paper we show that by training the two part DL model using the multi-environment meta learning scheme we observe improved positioning accuracy when transferring and fine-tuning to a new environment. In addition, the positioning performance of the models in each of the initial environments is not reduced. Our proposed approach enables simultaneous training of multiple models for different environments while further improves transfer learning to a new environment.

\section{Multi-environment Meta-Learning}
\label{sec:system model}

We consider an uplink setup with $N_R$ antennas at the BS and $N_T$ antennas at the UE. The UE transmits a reference signal on $N_C$ subcarriers within an orthogonal frequency division multiplexing (OFDM) symbol. The received uplink signal is used to estimate the CSI between the UE and the BS. The estimated channel over the $N_C$ subcarriers between the $k$-th receive antenna and the $m$-th transmit antenna can be described as: 

\begin{equation}
	\boldsymbol{h}^{k,m} = [{h}_0^{k,m}, {h}_1^{k,m}, ..., {h}_{N_C - 1}^{k,m}]^\text{T} \in \mathbb{C}^{N_C}.
\end{equation}
Furthermore, we define the measured channel $\tilde{\boldsymbol{H}}^m$ between the $m$-th transmit antenna of the UE and the BS as:

\begin{equation}
	\tilde{\boldsymbol{H}}^m =   [ \boldsymbol{h}^{1, m},   \boldsymbol{h}^{2, m},  ...,\boldsymbol{h}^{N_R, m} ] \in \mathbb{C}^{N_C \times N_R}.
\end{equation}
The CSI fingerprint is then obtained by stacking the $\tilde{\boldsymbol{H}}^m$ across the UE's transmit antennas:
\begin{equation}
	\tilde{\boldsymbol{H}} =  [\tilde{\boldsymbol{H}}^1, \tilde{\boldsymbol{H}}^2, ..., \tilde{\boldsymbol{H}}^{N_T}]^\text{T} \in \mathbb{C}^{N_A \times N_C},
\end{equation}
where $N_A = N_R \cdot N_T$. This is eventually processed according to the phase difference between antennas as described in \cite{Foliadis2021CSIBasedLW} by considering the $\tilde{\boldsymbol{H}}$ with $N_A$ antennas. This results in a 3-dimensional matrix, i.e. $\boldsymbol{H} \in \mathbb{R} ^{N_A \times N_C \times 3}$, where the third dimension includes the magnitude, the sine and cosine of the phase difference between antennas as separate stacked 2D matrices.

The matrix $\boldsymbol{H}$ is then considered as a fingerprint of the UE's position. For DL based positioning using CSI fingerprints, the input of the NN consists of the CSI fingerprint $\boldsymbol{H}$ of a measured uplink channel between a UE and the BS. The aim of the NN is to estimate the UE's positiong based on the CSI fingerprint, i.e. the ouput of the NN is the estimated UE's positon. Towards this end, a database is created that consists of processed CSI measurements for different UE's positions along with the respective UE's position label ${\boldsymbol{p}} \in \mathbb{R}^2$.


\subsection{Multi-environment Learning}
\label{sec:multi_scenario_learning}
It is evident that such training method would generally result in the model's parameters  $\epsilon$ being dependent on the environment, i.e. $f_\epsilon(\boldsymbol{H}) = \tilde{\boldsymbol{p}}$. This fact inhibits the models ability to clearly differentiate between environment independent feature extraction and environment dependent feature combination. Due to this, a possible transfer learning algorithm can't fully exploit environment independent knowledge that the model has acquired.

Our proposed multi-environment meta learning approach is shown in Fig. \ref{fig:multi-environment}. $N$ separate models are trained based on $N$ different source environments, where the $n$-th model $f_{\theta,\epsilon_n}(\boldsymbol{H}_n)$ is parameterized by the common parameters $\theta$ and the $n$-th environment specific parameters $\epsilon_n$ for  $n={1, 2, ..., N}$. We consider a function that extracts channel features $\boldsymbol{z}_n$ from the estimated fingerprint $\boldsymbol{H}_n$ of the $n$-th environment. By defining  $f_{\theta,\epsilon_n}(\boldsymbol{H}_n)=g_{\epsilon_n}(\phi_\theta(\boldsymbol{H}_n))$ and training the parameters $\theta$ of the initial layers $\phi_\theta(\cdot)$ on data from multiple environments we are forcing $\theta$  to be the same regardless of the environment and the models are encouraged to learn a common environment independent function $\phi_\theta(\cdot)$. The latter layers of the model $g_{\epsilon_n}(\cdot)$ combine the output of the $\phi_\theta(\cdot)$ function $\boldsymbol{z}_n$ in a way that embeds the distinct propagation information of the $n$-th environment , i.e. $\tilde{\boldsymbol{p}_n} = g_{\epsilon_n}(\boldsymbol{z}_n)$, since they are only trained on data from that particular environment. The channel features $\boldsymbol{z}_n$ are unknown and are implicitly learned by the DL-model. 

\subsection{Meta-Learning}
\label{sec:metalearning}
By training the DL model $\phi_\theta(\cdot)$ using the multi-environment learning scheme we essentially create a function which has already learned how it can extract relevant features from a new target environment as well. In other words, the model has already "learned how to learn". In the context of meta-learning, the purpose of multi-environment training is to learn a general purpose algorithm that can generalize across different source environments and enable each new target environment to be learned better \cite{Hospedales2021MetaLearningIN}. 

\begin{figure}[t!]
	\centerline{\includegraphics[scale=0.63]{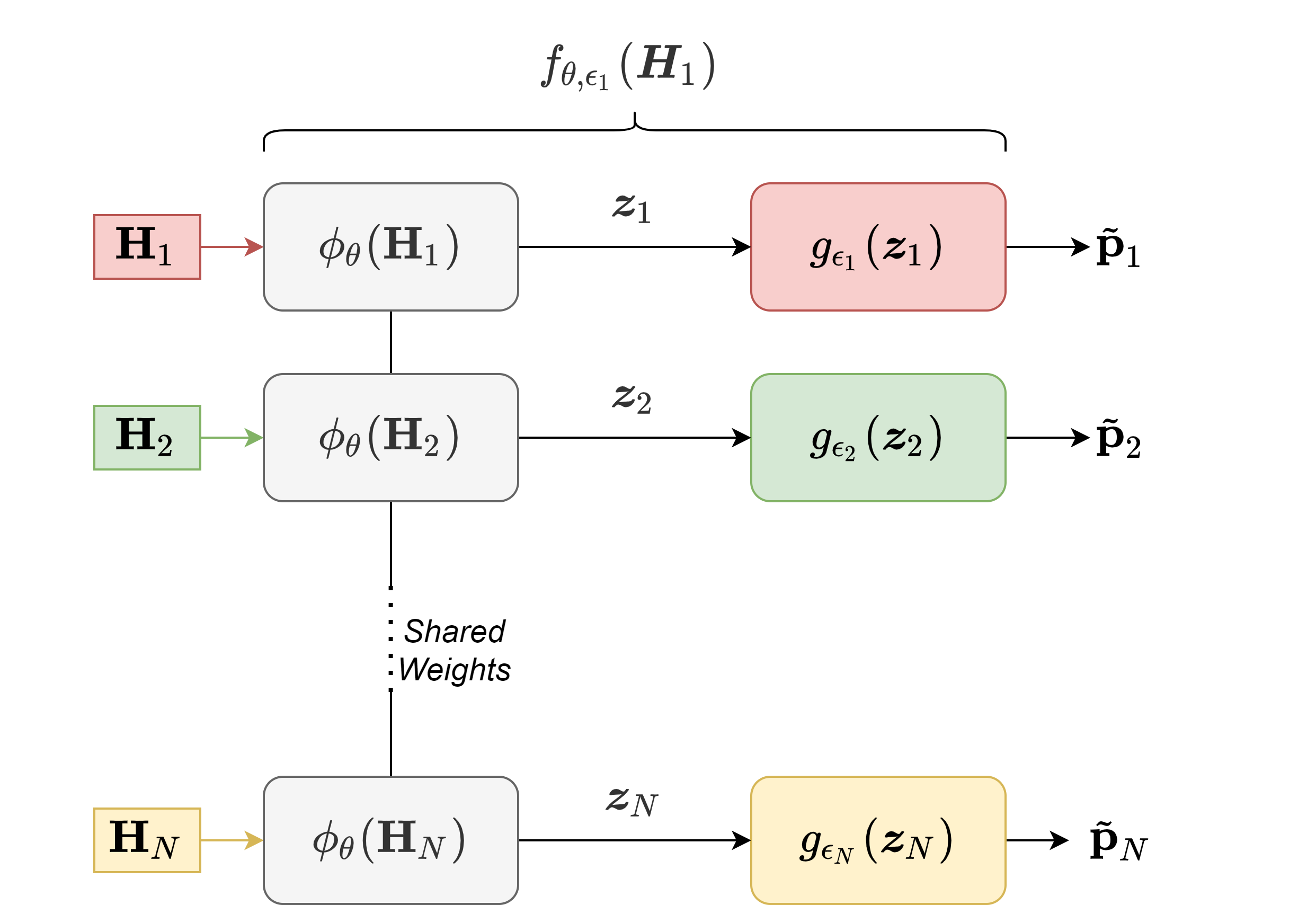}}
	\caption{Multi-environment learning over $N$ environments}
	\label{fig:multi-environment}
\end{figure}

\begin{figure}[t!]
	\centerline{\hspace{-0em}\includegraphics[scale=0.63]{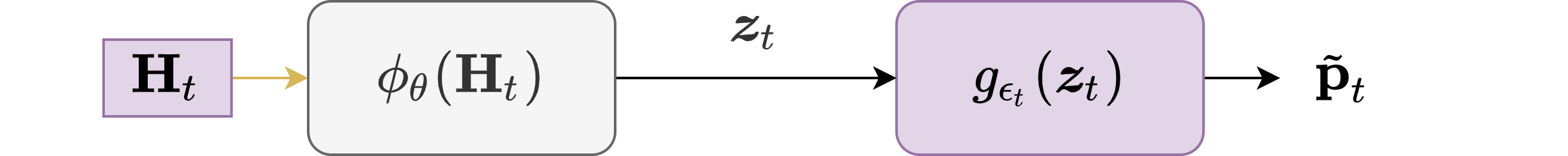}}
	\caption{Training of target environment using the knowledge from $\phi_\theta(\cdot)$}
	\label{fig:target-environment}
\end{figure}

The meta-objective that is optimized using the multi-environment meta-learning shown in Fig. \ref{fig:multi-environment} is as follows:

\begin{equation}
	\min_{\theta, \epsilon_n} \sum_n \mathcal{L}_n(f_{\theta, \epsilon_n}(\boldsymbol{H}_n)),
	\label{eq:metaobjective}
\end{equation}
where  $\mathcal{L}_n$ is the MSE loss of the $n$-th model for the $n$-th environment. By minimizing this objective when training the DL models, the weights of the layers $g_{\epsilon_n}(\cdot)$ are updated based only on the MSE loss $\mathcal{L}_n(f_{\theta, \epsilon_n}(\boldsymbol{H}_n))$ of the $n$-the model, while the common weights of $\phi_\theta(\cdot)$ are updated based on the summation of the individual losses \eqref{eq:metaobjective}. This effectively fulfills our intention of training the parameters $\theta$ on data from all $N$ environments and the parameters $\epsilon_n$ only from data from the $n$-th environment. The ultimate goal of the meta-learning scheme is to minimize the MSE loss on a new unseen environment $\epsilon_t$:

\begin{equation}
	\min_{\theta, \epsilon_t}  \mathcal{L}_t(f_{\theta, \epsilon_t}(\boldsymbol{H}_t)),
	\label{eq:objective}
\end{equation}
 We show that the learned $\phi_\theta(\cdot)$ provides good initial weights for the model $f_{\theta, \epsilon_t}(\cdot)$ when trained on the  new  test environment $t$ depicted in Fig. \ref{fig:target-environment}, enabling a more efficient minimization of the MSE loss for the new target environment. 

It is important to mention the relationship of our multi-environment meta-learning proposal to multi-task learning (MTL) algorithms \cite{Crawshaw2020MultiTaskLW} since the objective that is minimized in  \eqref{eq:metaobjective} is in itself a multi-task objective. The idea behind MTL is that by learning multiple tasks simultaneously while sharing parameters across the models of those different tasks, an implicit regularization is included in the learning process. Additionally, it can be considered as a data augmentation method since the training set size of the function $\phi_\theta(\cdot)$ is effectively increased compared to only having one environment. The difference between MTL and meta-learning though is that in meta-learning the goal is to acquire an algorithm that generalizes in unseen tasks (i.e. positioning in a new environment) by making minimization of the loss of the new task more efficient. On the other hand, with MTL, the objective is to learn effectively only the tasks at hand (i.e. positioning in the multiple $N$ environments). We show that our proposed multi-environment meta-learning scheme is able to achieve both of those goals.

\section{Simulation Setup}

\subsection{Database Description}
To evaluate our multi-environment meta-learning algorithm we used a ray-tracing dataset called DeepMIMO \cite{Alkhateeb2019DeepMIMOAG}. This dataset lets the user define different channel parameters and simulate a propagation environment from a UE at a particular position to several BSs in an outdoor urban environment. 

\begin{figure}[t]
	\centerline{\includegraphics[scale=0.235]{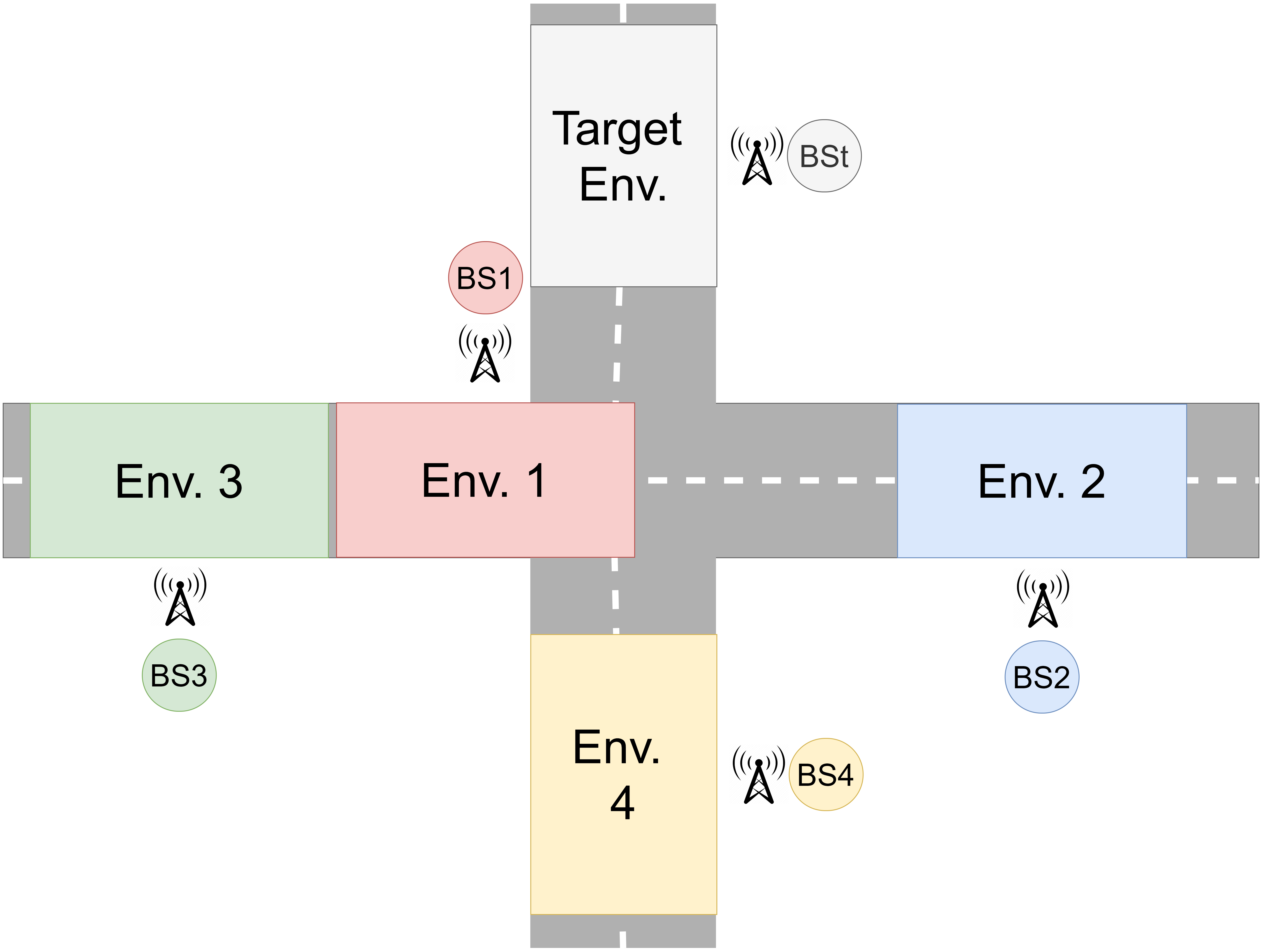}}
	\caption{Multiple considered environments of the DeepMIMO database}
	\label{fig:DeepMIMO}
\end{figure}

For our simulations we consider 5 different street segments each  one with one BS as depicted in Fig. 3. The area covered by each environment has the same dimensions with the same relative location of the BS, while the propagation paths for each environment differ because of the specific building configuration around each street segment. Out of those 5 environments, 4 serve as the source environments (Fig. \ref{fig:multi-environment}) for our proposed multi-environment meta-learning algorithm while the remaining one is the target environment (Fig. \ref{fig:target-environment}). Each BS has a uniform linear array (ULA) in the azimuth plane with $N_R = 8$ antennas that is always parallel to the street. The antenna array at the UE is a uniform rectangular array (URA) with $N_T = 4$ antennas. The spacing between adjacent antenna elements is half wavelength for both the UE and the BS. For this outdoor environment we consider a carrier frequency of $f_c = 3.5\text{GHz}$ and we assume a bandwidth of $BW = 100\text{MHz}$. The number of subcarriers is 1024 but we assume the reference signal is transmitted only on every 10th subcarrier which results in a CSI fingerprint with $N_C=52$ subcarriers. We also assume that the UE's transmit power is  23 $\text{dBm}$  per antenna, the noise floor is -174 $ \text{dBm/Hz}$ and the receiver's noise figure is 2 $\text{dB}$. For the training, we assume the same number of measurements is available from each of the source environments 1-4. In each environment we consider the positioning of a UE based on uplink channel measurements at the BS of that environment.

\subsection{Neural Network setup}

\begin{figure}[t]
	\centerline{\hspace{-2em}\includegraphics[scale=0.43]{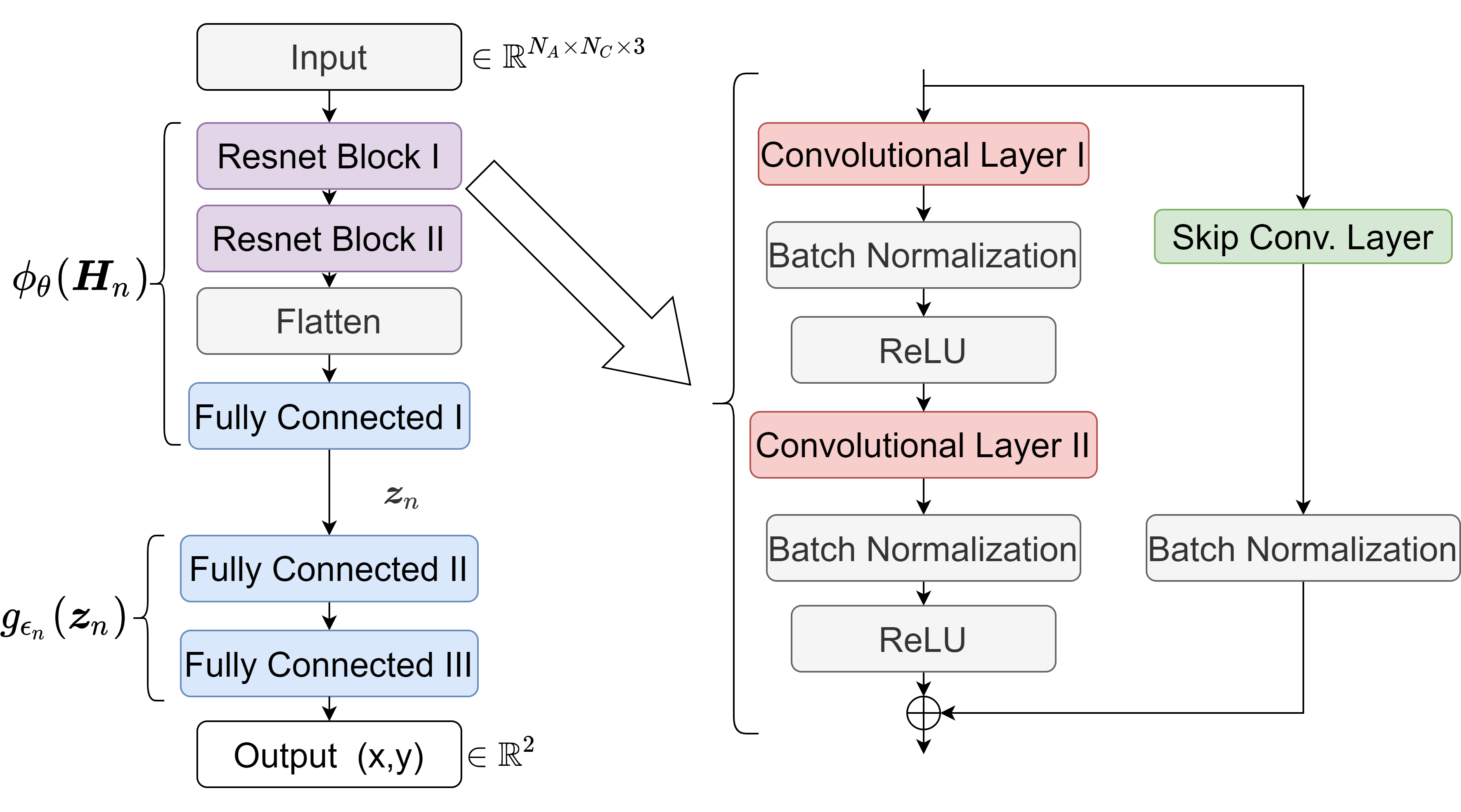}}
	\caption{Considered DL-model}
	\label{fig:neuralnetwork}
\end{figure}

For the DL-model we consider a convolutional neural network with residual layers \cite{He2016DeepRL}. As shown in Fig. \ref{fig:neuralnetwork}, the CSI fingerprint at the input first goes through two residual blocks. Each residual block consists of two convolutional layers and a skip connection. Each convolutional layer is followed by a batch normalization and a rectified linear unit (ReLU) activation function. No activation function is used for the skip connection. The batch normalization layers are used as a means to to mitigate the problem of internal covariate shift that affects the learning process of NNs \cite{Ioffe2015BatchNA}. The stride of the second convolutional layer in each residual block is $1 \times 4$, effectively downsampling the output in the subcarrier dimension only. The output of the residual blocks is flattened and inputted to three subsequent fully connected layers with 128 neurons each. Each of the convolutional layers has 32 filters and a $4 \times 4$ kernel.

As shown in Fig. \ref{fig:neuralnetwork}, the residual blocks and the first of the fully connected layers of the neural network are considered as the environment independent part $\phi_\theta(\cdot)$ that is jointly trained across multiple environments as described in section \ref{sec:multi_scenario_learning}. The remaining fully connected layers comprise the environment specific part $g_{\epsilon_n}(\boldsymbol{z}_n)$, that is trained only from data of the n-the environment. The number of weights corresponding to the function $\phi(\cdot)$ is $600k$ while for the  $g_{\epsilon_n}(\boldsymbol{z}_n)$ this number is $40k$. Thus, around $93.5\%$ of the weights are concentrated in the initial layers, which we aim to make environment independent. During training, a batch size of 64 was chosen and the number of epochs was set to 1000. All the inputs are normalized in the range $[0, 1]$. As an evaluation metric we consider the mean error (ME) of the UE positioning, which is given by the euclidean distance between the estimated position and the true position of the UE in the test set.

\section{Simulation Results}

We test the effectiveness of our multi-environment meta-learning approach by learning the function $\phi(\cdot)$ with data from the $N$ source environments and using it to initialize the layers of a new DL model which is trained on data from the target environment. We incrementally increase the number of source environments $N$ from $N=1$ to $N=4$. When $N=1$, this is equivalent to regular transfer learning with no multi-environment training scheme applied. Additionally for comparison, we train a new NN from scratch with only data from the target environment. The weights of the function $g_{\epsilon_t}(\boldsymbol{z}_t)$ are always initialized randomly for the target environment $t$. The number of $N_S$ training samples for each source environment is the same, while the target environment is trained for increasing training samples. We consider $N_S=180000$.

For the source environments, 80\% of the samples is used for training and 20\% for testing. For the target environment 30\% of the samples are used for testing. The other 70\% are potential training samples of the training set. We calculate the ME for different number of samples out of the test environment training set.  In the following the ME is calculated for test set. 

\begin{table}[b]
	\caption{ME with separate learning and multi-environment learning}
	\begin{center}
		\begin{tabular}{|c||c|c|c|c|}
			\hline
			&\textbf{Env. 1} & \textbf{Env. 2}& \textbf{Env. 3}& \textbf{Env. 4}\\
			\hline 
			Separately & 0.9785 m & 0.7242 m & 0.6766 m & 0.6532 m\\
			\hline 
			Jointly & 0.8442 m & 0.5697 m & 0.5368 m & 0.5815 m \\
			\hline 
		\end{tabular}
		\label{tab:mtl_results}
	\end{center}
\end{table}

As explained in section \ref{sec:metalearning}, training of the function $\phi(\cdot)$ and $g_{\epsilon_n}(\boldsymbol{z}_n)$ with data from the source environments is by itself a MTL problem. If we approach it from this perspective we expect the positioning accuracy even of the source environments to improve when applying the multi-environment meta learning scheme. In Table \ref{tab:mtl_results} we see the benefit of training the $\phi(\cdot)$ jointly on data from multiple environments, i.e. when the objective \eqref{eq:metaobjective} is minimized, compared to when it is trained separately for each environment, i.e. when only the MSE loss $\mathcal{L}_n(f_{\theta, \epsilon_n}(\boldsymbol{H}_n))$ is used to update both   $\phi(\cdot)$ and $g_{\epsilon_n}(\boldsymbol{z}_n)$. There is a clear decrease in the positioning error on the test set of each of the source environments. This means that besides transferring the knowledge to a target environment, our multi-environment meta learning scheme improves the positioning accuracy in each of the source environments as well.

\begin{figure}[tp]
	\hspace{-2em} \includegraphics[scale=0.65]{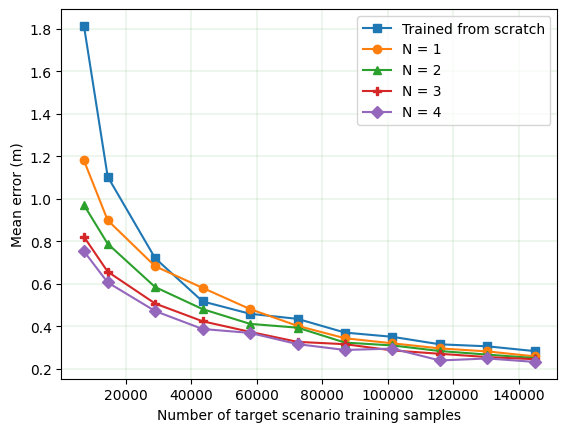}
	
	\caption{Mean error (ME) of multi environment meta learning scheme, with fine-tuning of $\phi(\cdot)$}
	\label{fig:error_nonfrozen}
\end{figure}

Next, the function $\phi(\cdot)$, which is used to initialize the weights of the respective layers for the new NN, is fine tuned with the data from the target environment. The results are shown in Fig. \ref{fig:error_nonfrozen}. We see that simple transfer learning ($N=1$) achieves better positioning performance than training the DL-model from scratch but only when the number of training samples from the target environment is small. The benefit disappears when the number of training samples is increased. However, this is not the case when applying our proposed multi-environment meta learning scheme, i.e. $N=[2, 3, 4]$. When increasing number of source environments we see a decreasing ME regardless of the target environment training samples. This indicates that the function $\phi(\cdot)$ learned with the multi-environment meta learning is indeed a good initialization for the fine tuning in the target environment. Thus, our proposed approach improves the transfer learning with increasing number of source environments.

\begin{figure}[b]
	\renewcommand{\thefigure}{7}
	\hspace{-2em} \includegraphics[scale=0.65]{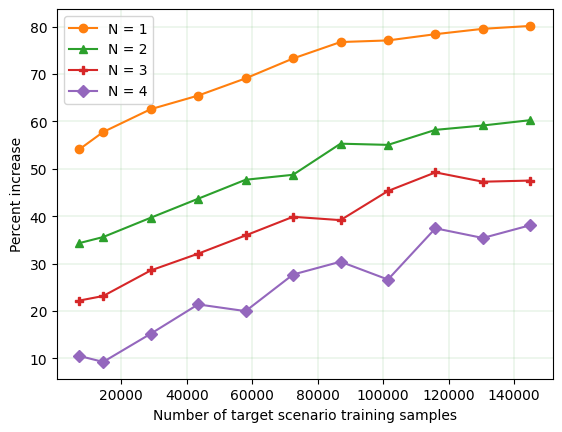}
	\caption{Mean error (ME) percent increase when freezing the initial layers}
	\label{fig:frozen_nonfrozen_comp}
\end{figure}

\begin{figure}[t]
	\renewcommand{\thefigure}{6}
	\hspace{-2em} \includegraphics[scale=0.65]{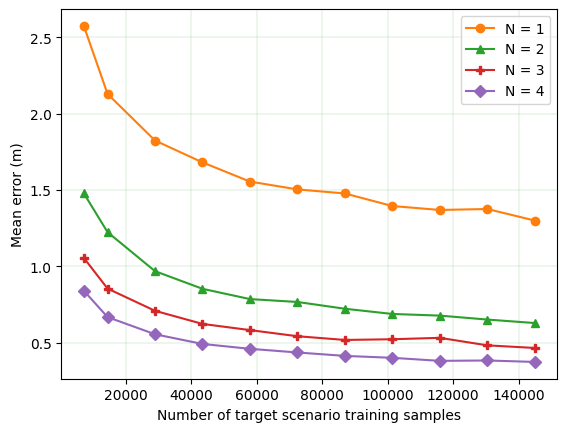}
	\caption{Mean error of multi environment meta learning scheme, with frozen $\phi(\cdot)$}
	\label{fig:error_frozen}
\end{figure}

We also examine the results when the function $\phi(\cdot)$ is frozen, in the DL model for the target environment, i.e., where only the function $g_{\epsilon_n}(\cdot)$ is trained with the data of the target environment $t$. This effectively means that the knowledge that $\phi(\cdot)$ embeds is only based on the source environments. The results for this case are shown in Fig. \ref{fig:error_frozen}. The case of trained from scratch is not included as it is not meaningful for the comparison, given that $\phi(\cdot)$ would be randomly initialized and thus, does not embed any knowledge for the positioning task. From Fig. \ref{fig:error_frozen} we can observe that the benefit of our multi-environment meta learning scheme is more pronounced. It is evident, that the function $\phi(\cdot)$ embeds knowledge that is helpful for the target environment, as by increasing the number of source environments, the ME on the target environment training set is decreased, by only needing to train the $g_{\epsilon_t}(\boldsymbol{z}_t)$ function with data from the target environment. 

Generally, freezing the layers of the NN results in worse performance but it may significantly decrease the training time. We now compare the two cases, i.e. when freezing and when fine-tuning the $\phi(\cdot)$ function in Fig. \ref{fig:frozen_nonfrozen_comp}. There we see the percentage increase of the ME when the layers corresponding to $\phi(\cdot)$ are frozen compared to when they are fine-tuned with CSI fingerprints from the target environment. As explained, we see that freezing the layers decreases the performance but the performance decrease becomes smaller when more source environments are used. We postulate that with enough source environments the performance decrease may be further reduced. When the performance difference between freezing and not freezing the layers is negligible, the benefit of lower training time of the DL model with frozen layers can be exploited without any downsides.

As discussed, one benefit of training the function $\phi(\cdot)$ on multiple environments is the implicit data augmentation that is happening during training with $N$ source environments. Assuming that the number of training samples for each environment is $N_S$, then the function $\phi(\cdot)$ is trained on $N \cdot N_S$ samples. In order to examine only the benefit of the knowledge obtained from multiple environments and not from training on more data, we employ only $Ns/N$ traninig samples for each source environment when considering $N$ source environments. This means that $\phi(\cdot)$ is always trained on $N_S$ samples regardless of the number of source environments.

\begin{figure}[!t]
	\hspace{-2em} \includegraphics[scale=0.65]{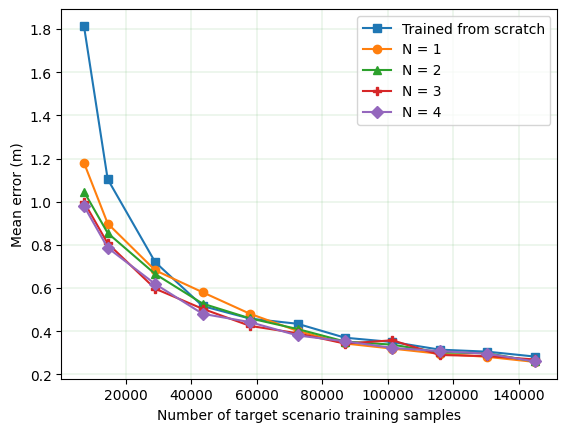}
	\caption{Constrained number of source  training samples, with fine-tuning of $\phi(\cdot)$}
	\label{fig:error_nonfrozen_divided}
\end{figure}

In Fig. \ref{fig:error_nonfrozen_divided} we see the effect of the knowledge that $\phi(\cdot)$ has gathered from training on multiple environments and not from the increased amount of training data. As expected, the transferring capabilities of the multi-environment meta learning scheme are reduced when the number of training samples of the $\phi(\cdot)$ is constrained. Regardless, there is still an advantage that is gained from training in multiple source environments, especially for lower number of training samples of the target environment. The saturation that occurs when going from $N=3$ to $N=4$ can be attributed to the fact that for these large urban areas, having $N_S / 4$ number of samples from each environment is not enough to fully capture the properties of each environment that aid the multi-environment meta-learning scheme. 

\begin{figure}[!t]
	\hspace{-2em} \includegraphics[scale=0.65]{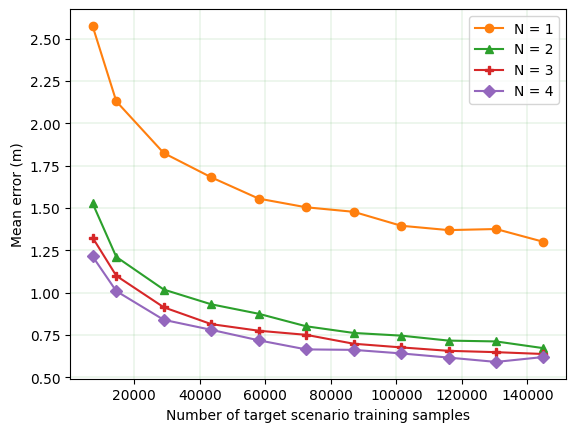}
	\caption{Constrained number of source  training samples, with frozen $\phi(\cdot)$ }
	\label{fig:error_frozen_divided}
\end{figure}

To check whether the initial layers $\phi(\cdot)$ have captured knowledge that can be beneficial to the target environment, we also consider freezing  $\phi(\cdot)$ with the constrained training for the multi-environment training, i.e, when training $\phi(\cdot)$ with $N_S/N$ samples for each of the $N$ source environments. The results are shown in Fig. \ref{fig:error_frozen_divided}. Indeed, for the case of constraining the training samples, the knowledge that $\phi(\cdot)$ has acquired is useful for the target environment as well.  Similar to Fig. \ref{fig:error_frozen}, the benefit of training  $\phi(\cdot)$ with multiple source environments is evident, even with reduced training samples per environment.

\section{Conclusion}

We propose a DL model consisting of two parts, where for the first part we aim to construct an environment independent function, while the second part combines features obtained from the first part with an environment dependent function. For this purpose, we propose a novel multi-environment meta learning scheme for training the first part based on multiple source environments and using this trained first part for improving the transfer learning in a new unseen target environment. For training in a target environment, we initialize the DL model based on the environment independent function learned over multiple source environments. We show that with our proposed approach, the positioning error in the target environment is reduced compared to regular transfer learning or compared to training the DL model of the target environment from scratch. The benefit is more pronounced with increasing number of source environments, thereby indicating that indeed the environment independent function is learning features that are applicable to unseen environments. The multi-environment learning is also beneficial when freezing the first part of the DL model when training in a new target enviroment or when constraining the number of training samples for the multi-environment training. With our proposed scheme, we are able to create a environment independent function which can embed knowledge from multiple environments and more effectively learn from a new environment.

\bibliographystyle{IEEEtran}
\bibliography{references}

\end{document}